\documentclass[journal]{IEEEtai}

\usepackage[colorlinks,urlcolor=blue,linkcolor=blue,citecolor=blue]{hyperref}

\usepackage{color,array}
\usepackage{threeparttable}

\usepackage{graphicx}

\begin{document}

\title{Wavelet-based Autoencoder and EfficientNet for Schizophrenia Detection from EEG Signals} 

\author{Umesh Kumar Naik M, and Shaik Rafi Ahamed, \IEEEmembership{Senior Member, IEEE}
\thanks{This paragraph of the first footnote will contain the date on which you submitted your paper for review. 
Umesh Kumar Naik M and Shaik Rafi Ahamed are with the Department of Electronics and Electrical Engineering, Indian Institute of Technology Guwahati, Assam, 781039, India, e-mail: (u.mudavath@iitg.ac.in and rafiahamed@iitg.ac.in)} }

\markboth{Journal of IEEE Transactions on Artificial Intelligence, Vol. 00, No. 0, Month 2020}
{Umesh Kumar Naik M. \MakeLowercase{\textit{et al.}}: Bare Demo of IEEEtai.cls for IEEE Journals of IEEE Transactions on Artificial Intelligence}

\maketitle

\begin{abstract}
Schizophrenia (SZ) is a complex mental disorder that necessitates accurate and timely diagnosis for effective treatment. Traditional methods for SZ classification often struggle to capture transient EEG features and face high computational complexity. This study proposes a convolutional autoencoder (CAE) to address these challenges by reducing dimensionality and computational complexity. Additionally, we introduce a novel approach utilizing spectral scalograms (SS) combined with EfficientNet (ENB) architectures. The SS, obtained through continuous wavelet transform, reveals temporal and spectral information of EEG signals, aiding in the identification of transient features. ENB models, through transfer learning (TL), extract discriminative features and improve SZ classification accuracy. Experimental evaluation on a comprehensive dataset demonstrates the efficacy of our approach, achieving a five-fold mean cross-validation accuracy of 98.5\% using CAE with a soft voting classifier and 99\% employing SS with the ENB7 model. These results suggest the potential of our methods to enhance SZ diagnosis, surpassing traditional deep learning (DL) and TL techniques. By leveraging CAE and ENBs, this research offers a robust framework for objective SZ classification, promoting early intervention and improved patient outcomes.

\end{abstract}

\begin{IEEEImpStatement}
This research significantly advances artificial intelligence (AI) in biomedical applications, particularly in detecting SZ using EEG data. The proposed models effectively address challenges related to efficacy, interpretability, and complexity, showcasing the transformative potential of AI in healthcare. This work presents novel data processing techniques for deriving insights from complicated biomedical data by integrating SS with ENB models and developing a CAE. The CAE model adeptly handles high-dimensional data, enhancing analysis accuracy and efficiency. Real-time health monitoring devices are made possible by the development and validation of these DL models, which transcend state-of-the-art approaches in terms of performance and adaptability.

\end{IEEEImpStatement}


\section{Introduction}



\IEEEPARstart{S}{chizophrenia} is a debilitating mental disorder that affects a significant portion of the global population. It is characterized by disturbances in thought processes, emotions, and behavior \cite{b1}, leading to profound social and occupational impairments. The World Health Organization (WHO) estimates that SZ affects about 24 million individuals worldwide \cite{b6}. Despite a century of research, the etiology of the disorder is still uncertain \cite{b7}, but there is substantial evidence of structural abnormalities associated with the disorder, which include ventricular enlargement, decreased cerebral volume (in cortical and hippocampal regions), and alterations in cerebral asymmetries \cite{b11}. EEG studies in schizophrenia have shown increased theta and delta waves \cite{b23}, decreased alpha waves \cite{b24}, and event-related potentials (ERP) abnormalities \cite{b25}. The aetiology of SZ still remains unclear, with a combination of genetic, environmental, and neuro-developmental factors believed to contribute to its onset. The timely and precise diagnosis of SZ is crucial for the efficient treatment and care of the illness. Timely intervention can significantly improve patient outcomes \cite{b26}, minimize symptom severity, and enhance overall quality of life. However, diagnosing SZ can be challenging due to the complexity and heterogeneity of its symptoms, often leading to delayed or missed diagnoses.

The impact of delayed diagnosis and treatment is substantial, as individuals with untreated SZ often experience worsening symptoms and functional decline over time or have not shown significant improvements despite being provided with pharmacological interventions \cite{b7}. Furthermore, the economic burden associated with untreated or poorly managed SZ is significant, encompassing healthcare costs, lost productivity, and social welfare expenditures. Moreover, childhood factors such as delayed motor development, speech problems, solitary play preference, low educational test scores, anxiety in social situations, and lower social confidence are linked to an elevated risk of adult SZ development \cite{b10}. While there is currently no cure for SZ, early diagnosis and appropriate treatment can help manage symptoms, reduce relapse rates, and improve long-term prognosis. Antipsychotic medications, psychotherapy, and psychosocial interventions form the cornerstone of treatment approaches. However, achieving optimal outcomes relies on accurate identification and prompt intervention, highlighting the critical role of early diagnosis in the management of SZ. 

Diagnosing SZ is challenging but crucial for effective treatment and management \cite{b21}. Finding SZ biomarkers and comprehending the electrical activity of the brain has been made possible with the use of EEG analysis. EEG recordings capture electrical signals generated by brain cells, offering insights into neural abnormalities associated with the disorder. By analysing specific patterns, frequencies, and connectivity measures in the EEG data, researchers can detect deviations that may indicate the presence of SZ. EEG analysis for SZ involves various approaches, including spectral analysis, ERPs, and connectivity analysis combined with various machine learning (ML) algorithms that aim to identify characteristic EEG signatures associated with the disorder, such as changes in the oscillatory rhythms \cite{b23} \cite{b24}, and abnormalities in ERPs \cite{b25} and functional connectivity patterns \cite{b27}. On a similar note, DL models have gained traction in SZ analysis using EEG data in recent years. These models leverage neural network technology to automatically learn complex patterns and make accurate predictions. They extract relevant features from the EEG signals and classify individuals as either healthy control (HC) or having SZ based on learned patterns.

In this research paper, we aim to address the difficulties in diagnosing SZ, such as the heterogeneous nature of the disorder, the intricacy of EEG patterns, and the problem of identifying discriminative features by proposing a novel approach that leverages discrete wavelet transform (DWT) combined with CAE along with SS combined with the family of ENB models \cite{b3}. By harnessing the potential of EEG signals and advanced DL techniques, our methodology seeks to improve the accuracy and efficiency of SZ classification, ultimately leading to early intervention and improved patient outcomes. The motivation behind research on SZ diagnosis using EEG lies in the potential to enhance early detection, improve treatment outcomes by effectively capturing transient EEG features, address high computational complexity, and advance our understanding of the neurobiology underlying the SZ disorder. EEG-based diagnostic approaches offer a promising avenue for objective and efficient assessment, paving the way for more effective interventions and improved quality of life for individuals living with SZ. 

\section{RELATED WORKS}

\begin{figure*}[!h]
\centering
\centerline{\includegraphics[width=\linewidth]{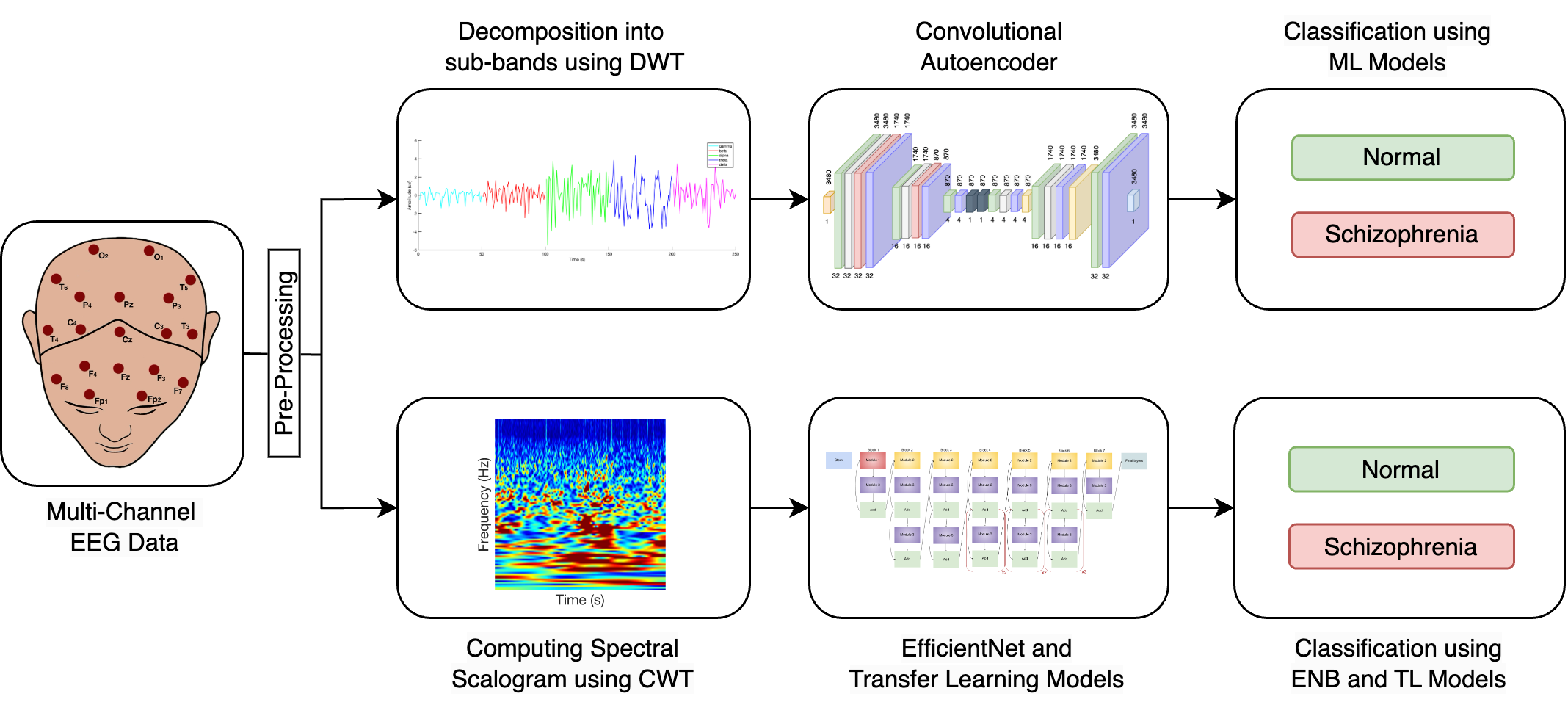}}
\caption{General pipeline of the proposed model framework}
\label{fig}
\end{figure*}

Researchers have employed different approaches to understand the underlying mechanisms, diagnose the condition, and develop effective SZ treatment strategies. In recent years, there has been a growing interest in utilizing ML and DL approaches for SZ classification, with a particular focus on analysing EEG signals. An often-used strategy in research involves exploring different hand-crafted feature (HCF) and latent features \cite{b22} extraction methods and employing classification algorithms to discriminate between individuals with SZ and HC. Some commonly used HCFs include statistical measures (first and second order) and time-frequency attributes (Shannon entropy, dominant frequency, spike rhythmicity, etc.) \cite{b5}. For instance, Yang et al. \cite{b12} developed a hybrid ML method combining functional magnetic resonance imaging (fMRI) and single nucleotide polymorphism (SNP) data for the classification of SZ. The proposed method involved selecting discriminating SNPs and fMRI voxels to construct separate support vector machine (SVM) ensembles, as well as utilizing independent component analysis to extract fMRI activation components. These models were combined using a majority voting approach, resulting in improved classification accuracy compared to using genetic or fMRI data alone. The study demonstrates the potential of integrating genetic and brain function data to better understand the etiopathology of SZ and identify diagnostically important markers for the disorder. 

As standard univariate analysis of neuroimaging data has limitations in clinical translation, Orru et al. \cite{b13} explored alternative analysis by doing a thorough review employing SVM, which has shown success in disease diagnosis, transition prediction, and treatment prognosis using functional and structural neuroimaging data and claimed despite posing theoretical and practical challenges it holds potential for future advancements in clinical settings for the identification of imaging biomarkers. Johannesen et al. \cite{b14} specifically employed the 1-norm SVM method to extract features from EEG data. The study focused on a Sternberg Working Memory Task (SWMT) and analyzed EEG data across different frequency components, processing stages, and scalp sites. The SVM models successfully classified SWMT trial accuracy and diagnosis, with frontal gamma and central theta being identified as primary classifiers. The results align with existing literature on the role of gamma and theta in memory processes and the presence of abnormal resting low-frequency activity in schizophrenia. 

Oh et al. \cite{b15} developed a computer-aided diagnosis (CAD) system utilizing an eleven-layered convolutional neural network (CNN) for the diagnosis of SZ, which automatically extracts significant features and classifies them from the EEG data. Similarly, Jahmunah et al. \cite{b16} aimed to develop a CAD tool to classify EEG signal patterns into HC and SZ. This approach utilized non-linear feature extraction, feature selection, classification, and validation. The SVM with radial basis function (SVM-RBF) achieved the highest average performance of 92.91\% in classification. Singh et al. \cite{b17} proposed a spectral features-based CNN (SFC) model to identify SZ patients from multichannel EEG signals. The model utilizes spectral analysis, dividing signals into distinct spectral bands and extracting spectral features, which are fed into the CNN and long short-term memory network (LSTM) models in order to classify data. The SFC model demonstrates high classification accuracies of 94.08\% and 98.56\%, respectively, for different datasets, showing promise for real-time and accurate diagnosis of SZ using EEG signals. Jindal et al. \cite{b18} introduced an automated CAD system for the detection of SZ using multichannel EEG activity. The methodology involves using a multisynchrosqueezing transform (MSST) to analyse the time-frequency representation of EEG signals. A Bi-CNN model is used to classify the decomposed EEG after features have been extracted. For SZ detection, the MSST-Bi-CNN approach obtained an accuracy of 84.42\%. Furthermore, Siuly et al. \cite{b19} proposed "SchizoGoogLeNet," a DL-based feature extraction scheme using the GoogLeNet model, for effective and automated detection of SZ from EEG signals. The proposed framework involves denoising the EEG signals and extracting deep features using the GoogLeNet model. Several ML classifiers, including GoogLeNet, are used to assess the deep feature set and obtain the highest accuracy of 98.84\%.

Current automatic classification approaches are based on very facile learning models that are insufficient for learning complicated and non-stationary EEG data. Other techniques may not effectively capture this multi-resolution decomposition, making SS particularly suited for revealing transient features and spectral characteristics in EEG signals. To address these issues, we propose a framework for the classification of SZ from the EEG data using CAE and ENBs. The extraction of time-frequency representations has garnered significant interest in the field of SZ EEG analysis. Wavelet-based approaches, using techniques such as continuous wavelet transform (CWT) and DWT, emerged as promising methods to capture the temporal and spectral characteristics of EEG data \cite{b29}. By providing a comprehensive visualization of the signal's frequency content over time, these approaches facilitate the identification of specific patterns or abnormalities within the EEG data. Furthermore, they provide detailed frequency resolution, quantifying power within specific EEG bands. These spectral features serve as potential biomarkers, aiding clinical interpretation and providing insights into neurophysiological mechanisms underlying cognitive processes. 

Figure 1 illustrates the comprehensive approach used for classifying SZ from EEG data. Initially, the pre-processed multi-channel EEG data was decomposed into sub-bands employing DWT, which reduces redundancy compared to CWT and facilitates a more efficient analysis. The resulting sub-band signals are fed into a CAE, which provides a sparse representation and mitigates computational complexity, allowing for real-time processing of high-dimensional EEG data. Simultaneously, SS are computed using CWT to capture both temporal and spectral features of the EEG signals, providing a rich representation through a multiscale analysis. This process enhances the discriminatory power of the classification model by detecting subtle spectral changes indicative of SZ \cite{b31}. The extracted SS are trained using a series ENB and a few other TL models. ENB's deep architecture facilitates automatic feature extraction from the SS, reducing reliance on manual feature engineering \cite{b3}. This enables the model to learn hierarchical representations and discern complex patterns, effectively discriminating between SZ and HC individuals \cite{b32}. TL leverages pre-trained knowledge from large-scale datasets, allowing the model to generalize well on smaller, domain-specific datasets, such as those available for SZ research. This approach alleviates the need for large amounts of labelled data, which is often a limitation in biomedical studies. Finally, we performed classification employing various ML models, resulting in the identification of individuals as either HC or SZ.

\section{METHODOLGY}

\subsection{Dataset description}
The dataset \cite{b2} used in this study focuses on adolescents in the age range
of 10 to 14. 45 boys, ages 11 to 14, who have been diagnosed with SZ make up the first group; 39 boys, ages 10 to 13, who are in the healthy group, make up the second. EEG segments were collected from awake adolescents in a relaxed state, with their eyes closed. To ensure the integrity of the EEG results, the subjects had not taken any medication prior to the study. The EEG signals were recorded using a 16-electrode system with reference to the 10-20 system. The electrode locations include 'F7', 'F3', 'F4', 'F8', 'T3', 'C3', 'Cz', 'C4', 'T4', 'T5', 'P3', 'Pz', 'P4', 'T6', 'O1', and 'O2'. The impedance of the electrodes was below 10 kOhm, the sampling rate was 128 Hz, and the bandwidth was set between 0.5 and 45 Hz.

The dataset underwent careful artifact removal, with both external and internal artifacts considered. External artifacts result from external influences, while internal artifacts are associated with subject-related actions such as muscular, ocular, and cardiac artifacts. Each EEG recording lasted for 60 seconds, resulting in a total of 7,680 samples for each channel. 

\begin{figure*}[h!]
\centering
\centerline{\includegraphics[width=\linewidth]{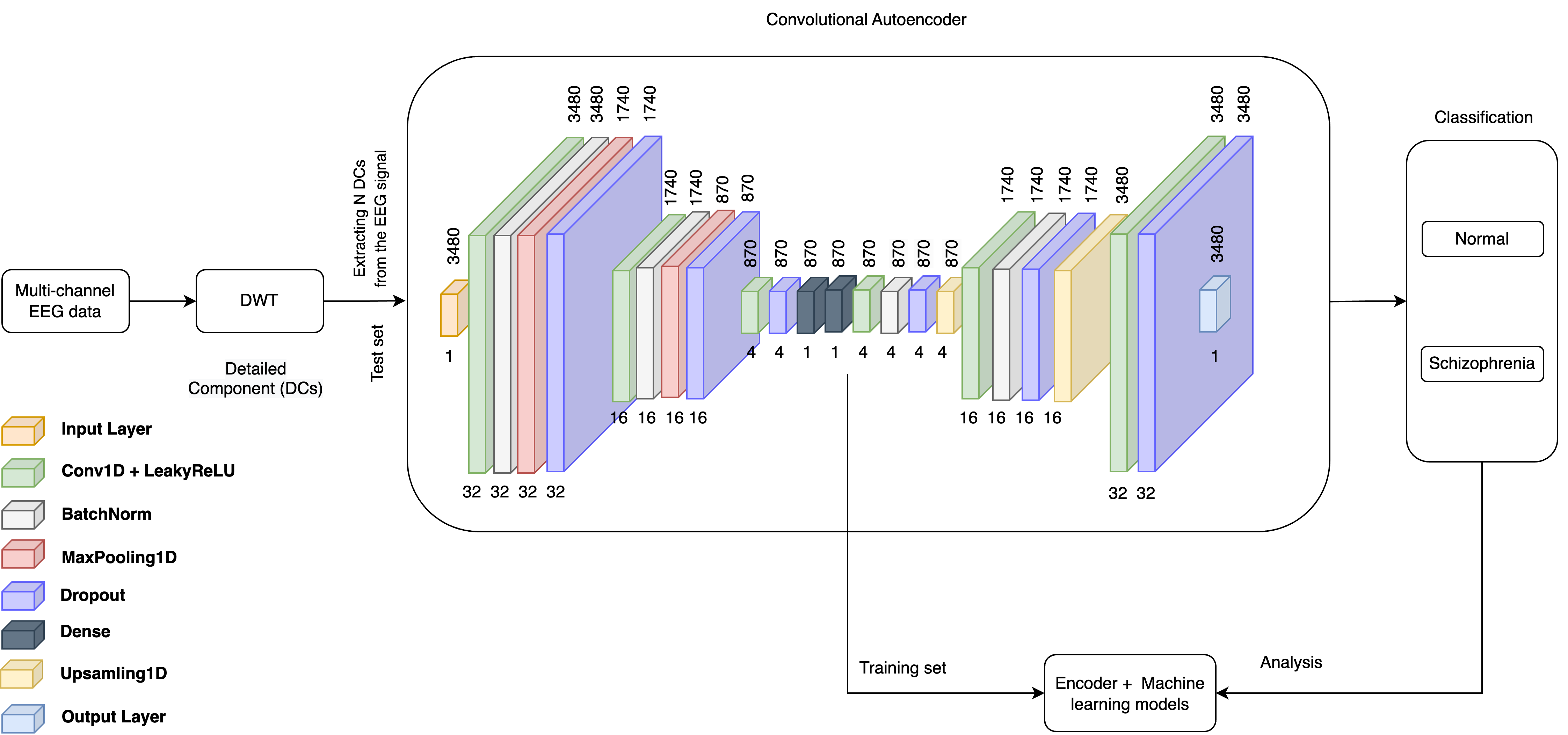}}
\caption{The system architecture of the proposed convolutional autoencoder model framework}
\label{fig}
\end{figure*}

\subsection{DWT for decomposition}

We employed DWT \cite{b9} instead of a bandpass filter (BPF) due to several advantages. DWT offers superior multiresolution capability, enabling signal decomposition into different frequency bands at various resolutions. It also provides better time-frequency localization and is more effective at reducing noise in signals compared to BPF. And is particularly suitable for evaluating non-stationary signals like EEG. By segmenting the EEG signal into diverse frequency bands followed by downsampling, DWT allows for efficient analysis with reduced time complexity. The EEG signal is decomposed into finer details by employing the following equations \cite{b5}:

\begin{equation}
W(t)=\sum_{l \in Z} 2^{\frac{n}{2}} a_{n}(k) \varphi\left(2^{n} t-k\right)+\sum_{n=0}^{n-1} \sum_{k=0}^{\infty} 2^{\frac{n}{2}} d_{n}(k) \psi\left(2^{n} t-k\right)
\end{equation}

\begin{equation}
a_{n}(k)=\int_{-\infty}^{\infty} W(t) \varphi\left(2^{n} t-k\right) d t
\end{equation}

\begin{equation}
d_{n}(k)=\int_{-\infty}^{\infty} W(t) \psi\left(2^{n} t-k\right) d t
\end{equation}
where $\psi(t)$ mother wavelet, and $\varphi(t)$  is a basic scaling function. In (2) \& (3), $a_n$ and $d_n$ are approximate and detailed coefficients of the segmented EEG signal, respectively.

\subsection{Convolutional autoencoder (CAE)}

The CAE comprises an encoder and decoder network, which is trained to learn the compact hierarchical representations of the input EEG data, which are crucial for differentiating between HC individuals and those with SZ. Specifically, we employ a wavelet-based CAE, which employs DWT to pre-process the EEG data. The CAE extracts the multivariate features from the bottleneck layer, which are used as input to various ML models for the classification of SZ.  Fig. 2 depicts the detailed system architecture of the CAE model framework. Overall, the CAE offers a data-driven and efficient approach for SZ classification from EEG, aiding in the accurate diagnosis and understanding of this complex neurological disorder.

Mathematically, let \textbf{$i = \mathbf{I}_{initial}^{S\times H\times W\times1}$} represent the input EEG data, where initial W and H are the number of amplitudes corresponding to the Temporal width and Height of each EEG signal $\mathbf{I}$, and $S$ is the number of samples. Moreover, E and D represent the Encoder and Decoder of the CAE. The representation of the encoder is given by, 
$j = \mathbf{Z}_{compressed}^{S\times d\times w\times1}$, where d and w are the depth and width dimensions of the compressed temporal 1D representation, and the representation of the decoder reconstruction is given by, 
$\hat{i} = \mathbf{I}_{initial}^{S\times H\times W\times1}$ comprises the encoder and decoder outputs, respectively, and is illustrated as:
\begin{equation}
\left\{\begin{array}{l}
j=E(i) \\
\hat{i}=D(j)
\end{array}\right.
\end{equation}

The reconstruction error $r^{wDAE}$ can be used to evaluate the CAEs' performance, which is given by:
\begin{equation} 
r^{CAE} = M_{CAE}{(\mathbf{\hat{z}}^{(x)},\mathbf{z}^{(x)})}
\end{equation}

The $M_{CAE}$ function represents a measurement of difference much like the ubiquitous square Euclidean distance is as follows:
\begin{equation} 
M_{CAE}(\mathbf{\hat{z}}^{(x)},\mathbf{z}^{(x)}) = \frac {1}{2} \| {\mathbf{\hat{z}}^{(x)} - \mathbf{z}^{(x)}} \|^2
\end{equation}

The cost function can therefore be expressed in its generic form as follows:
\begin{equation} \label{eq4}
N_{CAE} = \frac {1}{S} \sum_{x=1}^{S} L_{CAE}(D(E(\mathbf{z} )^{(x)})),\mathbf{z}^{(x)})
\end{equation}

Our goal is to minimise the cost function $N_{CAE}$ in order to determine the ideal weight parameters for the CAE. In our implementation, The cost function was minimised by using the Adam optimisation technique, with the learning rate set to ${10^{- 3}}$. To address the "dying ReLU" problem \cite{b4}, we chose LeakyReLU as the activation function, enabling gradient propagation through negative values. A negative slope parameter (alpha) of 0.2 was selected to balance gradient flow and preserve non-linearity.

\subsection{Scalogram}
A CWT is a mathematical technique used to analyse non-stationary signals, like EEG signals, in the context of SZ.  By computing SS using a continuous wavelet filter bank, the distribution of energy in the EEG signal can be represented over time and frequency. This allows us to identify complex patterns of activity that may be associated with specific brain processes or states relevant to SZ. SS analysis is a valuable tool in our work; as we apply it to EEG signals, it becomes possible to reveal hidden patterns of brain activity that may not be evident in the raw data alone. This is particularly significant in SZ analysis as the disorder involves complex disruptions in neural activity. They offer a holistic view of both temporal and spectral information, which helps to distinguish different EEG patterns linked to SZ. 

We chose SS (using CWT) over spectrograms (using Short-Time Fourier Transform (STFT)) \cite{b8} due to their superior resolution, better feature extraction capabilities, and ability to analyse non-stationary signals, which provide a detailed view at different times and frequency scales. A wavelet is characterised by its oscillatory behavior, finite energy, and zero average function. The CWT representation of a signal s(t) is mathematically denoted as follows \cite{b20}:

\begin{equation}
C_{a, b}[s(t)]=\frac{1}{\sqrt{a}} \int_{-\infty}^{\infty} s(t) \psi^{*}\left(\frac{t-b}{a}\right)
\end{equation}

In the wavelet transform equation, the scaling parameter $a$ determines the width of the wavelet function, while $b$ represents the time shift. The term $\psi(\frac{t-b}{a})$ corresponds to the mother wavelet $\psi(t)$ that has been shifted and scaled accordingly. The key advantage of the CWT is its flexibility in adjusting the window size. Low-frequency components can be analysed with a larger window, while high-frequency components can be analysed with a smaller window. This adaptability enables the CWT to effectively capture different scales of information in the signal.

\subsection{EfficientNets}

In recent times, ENB models \cite{b3} have gained superiority over conventional DL models due to their unique architectural design and compound scaling. Their architecture combines depth-wise separable convolutions, squeeze-and-excitation blocks, and inverted residual connections, enabling them to effectively capture complex features. Additionally, compound scaling optimizes the balance between model depth, width, and resolution, resulting in computationally efficient yet highly accurate models \cite{b3}. The resolution refers to the input image size, which affects the level of detail that the model can perceive. Efficient resource utilization is another strength of the ENB models \cite{b3}, as they may attain state-of-the-art results with fewer parameters and computational resources. This efficiency reduces memory usage, accelerates training and inference, and facilitates deployment on resource-constrained devices. ENB models also benefit from TL by making use of pre-trained weights from large-scale image datasets, allowing them to quickly adapt to new tasks with smaller labelled datasets. Finally, the generalization capability of ENB models is exceptional \cite{b3}, making them suitable for various image recognition tasks, including biomedical image analysis. Their architecture design and scaling factors enable them to learn transferable and discriminative features that generalize well across different domains.

In this work, we employed a series of ENB models that have been optimized for achieving better accuracy with minimal resource utilization, ranging from ENB0 to ENB7. ENB0 serves as the baseline model, with moderate depth and width. As we move from ENB0 to ENB7, the models progressively increase in both depth and width, resulting in larger and more powerful architectures. ENB7 represents the most advanced and complex model in the series. 

The architecture search process employed in ENB involves discovering the optimal network architecture that balances model size and performance. Through this process, the models are designed to be highly efficient, achieving state-of-the-art results with fewer parameters and computational resources compared to other models \cite{b3} \cite{b53}. When using the ENB models for SZ classification, we fine-tuned them on our specific dataset. A pre-trained ENB model—typically trained on large-scale image datasets like ImageNet—must be fine-tuned to our intended objective. By using the TL approach, the model is able to effectively extract the features and classify them by utilising the learned representations from the pre-training phase. Moreover, we adjusted the final classification layer of the ENB models to match the number of classes in our task, which in this case is two (SZ and HC). The loss function that we used was binary cross-entropy, and the model was trained using the Adam optimizer. 

\section{EXPERIMENTAL RESULTS}
\subsection{Pre-processing of the data}
 Initially, the EEG data of all the subjects were normalized to remove any inherent scaling differences and ensure that all data points were on a consistent and comparable scale for accurate analysis and interpretation. The actual length of the EEG data for one subject is 122880, which was reshaped to 16x7680 as the first 7680 samples represent 1st channel,  the second 7680 samples represent 2nd channel, and so on as per the dataset description \cite{b2}. For the CAE approach, the EEG signals were decomposed into various frequency bands (delta, theta, alpha, beta, and gamma \cite{b5}) and concatenated before feeding to the CAE model for training purposes. 
 
 For the ENB and TL model approach, by utilizing a CWT filter bank, we transformed the time series EEG data shown into informative 2D scalogram images. The CWT filter bank was constructed using the 'bump' wavelet with a total of 12 voices per octave, the sampling frequency of the EEG data is set to 128 Hz, and the frequency range for analysis is set between 0.5 Hz and 50 Hz. By setting these parameters, the CWT filter bank was efficiently analysing the EEG signal in the specified frequency range. Finally, these SS were fed to a family of ENB models for the classification of EEG data into SZ or HC. In order to compare and contrast the performance of SS, we also applied the ENB and TL models to spectrograms generated using STFT. For the STFT spectrogram, we used a rectangular window of length 256, with 50\% overlap, and 512 FFT points. Please refer to Appendix A of the supplementary material for the CWT scalograms and STFT spectrogram plots of the EEG signal. An Apple MacBook Pro equipped with an M1 processor is used to implement the proposed approach on MATLAB 2021a for pre-processing and Python 3.8 with Google Colab for ML and DL model development and deployment. 

\subsection{Machine Learning and Deep Learning}

The implementation of the models involved several steps. Firstly, the dataset was annotated and then shuffled to guarantee the independence of each data point and eliminate any potential bias. For the CAE, the encoder output of CAE after training was fed as input for various ML models, including random forest (RF), extreme gradient boosting (XGB), support vector classifier (SVC), k-nearest neighbors (KNN), and soft voting classifier (VC) classifier, which were employed for classification purposes. The classification data were split into training (68\%), validation (12\%), and testing (20\%) sets. The aforementioned models were trained using the training and validation data and evaluated by computing performance measures using the testing data with five-fold cross-validation. To ensure fair comparisons, we conducted experiments individually using each model, evaluating their performance in terms of various classification measures such as accuracy, recall, precision, F1-score, area under receiver operating characteristic curve (AUC), and Cohen's Kappa coefficient (KAP). This allowed us to assess the impact of model depth, width, and resolution on the classification task and identify the most suitable model for SZ classification. 

Subsequently, for the ENB models, the SS was utilized as input for a series of ENB models (ENB0-ENB7) along with six other TL models such as InceptionV3 \cite{b55}, VGG16 \cite{b56}, VGG19\cite{b56}, ResNet50 \cite{b56}, Xception \cite{b57}, and ConvNeXtSmall \cite{b58} were employed for classification purposes. Along with the aforementioned models, we also developed a 2D-CNN architecture comprising four convolutional layers alternated with four max-pooling layers. This was followed by four dense layers. The model was compiled using the Adam optimizer and employed binary cross-entropy loss. By employing the different ML and DL approaches, we aimed to explore their varying capabilities in capturing discriminative features from EEG data and achieving the highest possible accuracy during SZ classification. 



\subsection{Classification Results}

Based on the SS retrieved from the EEG data, including the eight ENB models, an additional six TL, and 2D-CNN models were implemented for the classification between SZ and HC. The target value is either SZ or HC (where SZ is considered 0, and HC is considered 1). The classification performance has been assessed based on various measures, i.e., accuracy, precision, recall, KAP, and AUC. Tables I and II depict the performance metrics for the eight ENB and six TL models combined with SS and STFT, respectively. Table III depicts the performance metrics for the CAE combined with ML classifiers. It is noteworthy that the CAE had a total of 38,296 trainable parameters and 136 non-trainable parameters, which are significantly fewer than those of the ENB or TL models. These results indicate that simpler models, which have fewer parameters and lower computational requirements, can perform on par with, and sometimes even surpass, more complex models. Fig. 3 depicts the confusion matrix of the CAE model with VC. To further validate and visualize the performance, the AUC curve for all the ENB and ML models has been illustrated in Fig. 4 and Fig. 5. Please refer to Appendix B of the supplementary material for the confusion matrix plots of all eight ENB models, and four other confusion matrix plots for CAE combined with ML classifiers. 

\begin{figure}[b!] 
\centering
\centerline{\includegraphics[width=\linewidth]
{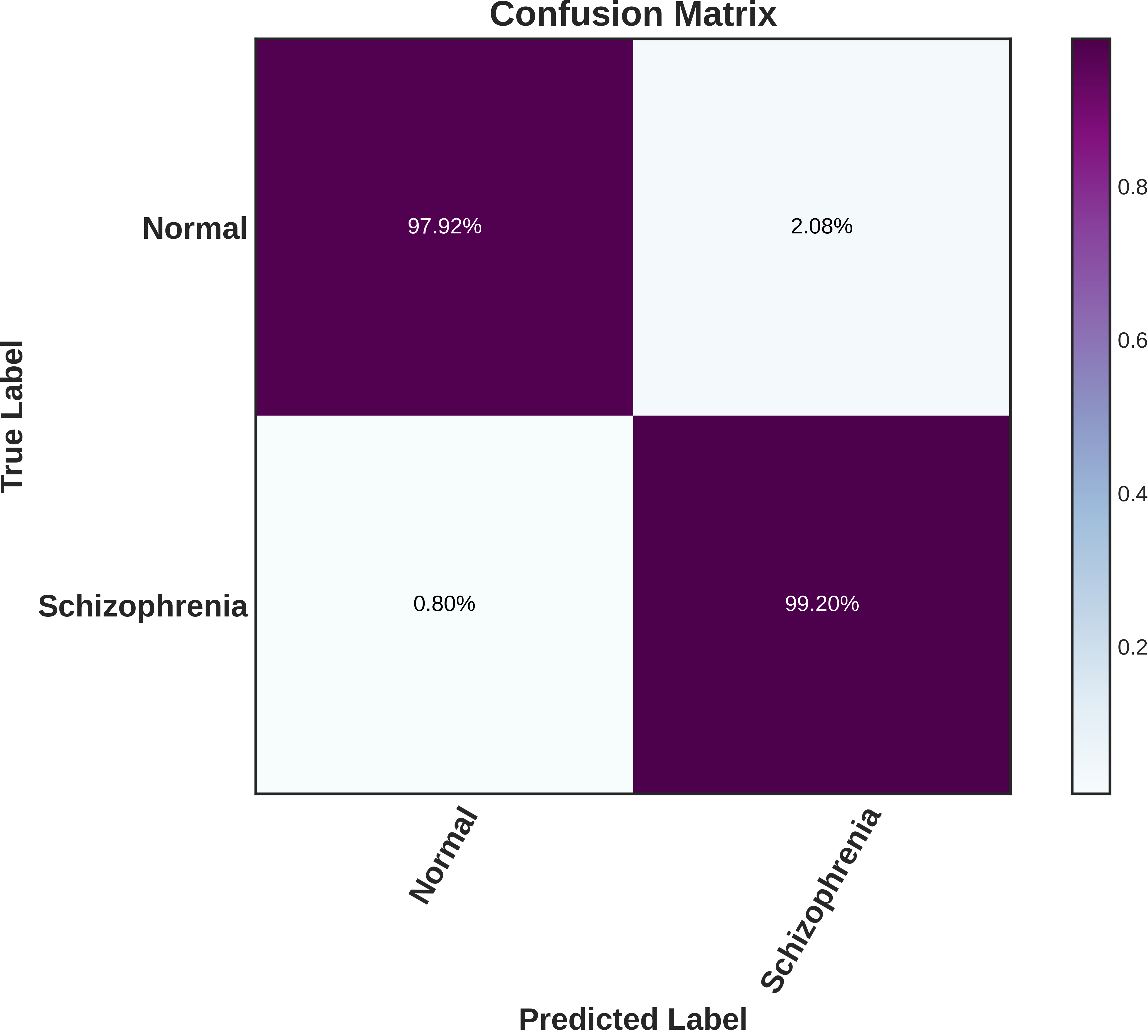}}
\caption{Confusion matrix of CAE with VC model}
\label{fig}
\end{figure} 


\begin{table*}[t]
\centering
\caption {Performance metrics for all the ENB models combined with SS and STFT} 
\begin{tabular}{l|lllll|lllll|l|l}
\hline \hline
  \textbf{Models}  & \multicolumn{5}{c|}{\textbf{SS}} &  \multicolumn{5}{c|}{\textbf{STFT}}   & \textbf{TP} & \textbf{NTP} \\ 
\cline{2-5} \cline{6-11} 
       &  \textbf{ACC}  & \textbf{PRE} & \textbf{REC} & \textbf{AUC} & \textbf{KAP}                                                             
	 &  \textbf{ACC} & \textbf{PRE}  & \textbf{REC} & \textbf{AUC} & \textbf{KAP} & & \\    \hline
    EfficientNet-B0 & 0.96 & 0.96 & 0.96 & 0.998 & 0.918  & 0.91 & 0.91 & 0.91 & 0.970 & 0.820 & 4,010,110 & 42,023 \\\hline
    EfficientNet-B1 & 0.94 & 0.95 & 0.94 & 0.997 & 0.881  & 0.95 & 0.95 & 0.95 & 0.989 & 0.903 & 6,515,746 & 62,055 \\\hline
    EfficientNet-B2 & 0.96 & 0.96 & 0.96 & 0.994 & 0.925 & 0.94 & 0.94 & 0.94 & 0.989 & 0.880  & 7,703,812 & 67,575 \\\hline
    EfficientNet-B3 & 0.97 & 0.97 & 0.97 & 0.997 & 0.932  & 0.91 & 0.92 & 0.91 & 0.987 & 0.826 &  10,699,306 & 87,303 \\\hline
    EfficientNet-B4 & 0.92 & 0.92 & 0.92 & 0.971 & 0.842  & 0.96 & 0.96 & 0.96 & 0.994 & 0.925 & 17,552,202 & 125,207 \\\hline
    EfficientNet-B5 & 0.96 & 0.96 & 0.96 & 0.994 & 0.917  & 0.97 & 0.97 & 0.97 & 0.995 & 0.940 & 28,344,882 & 172,743 \\\hline
    EfficientNet-B6 & 0.96 & 0.96 & 0.96 & 0.997 & 0.925  & 0.96 & 0.96 & 0.96 & 0.997 & 0.910 & 40,740,314 & 224,439 \\\hline
    EfficientNet-B7 & 0.99 & 0.99 & 0.99 & 0.999 & 0.985  & 0.97 & 0.97 & 0.97 & 0.996 & 0.932 & 63,792,082 & 310,727 \\
	  \hline \hline
\end{tabular}
  \begin{tablenotes}

            \item ACC - Accuracy, PRE - Precision, REC - Recall, KAP - Cohen's Kappa coefficient, TP - Trainable parameters, NTP - Non-trainable parameters
                    
 \end{tablenotes}
\end{table*}


\begin{table*}[t]
\centering
\caption {Performance metrics for other TL models compared with 2D-CNN and ENB0 models} 
\begin{tabular}{l|lllll|lllll|l|l}
\hline \hline
  \textbf{Models}  & \multicolumn{5}{c|}{\textbf{SS}} &  \multicolumn{5}{c|}{\textbf{STFT}}   & \textbf{TP} & \textbf{NTP} \\ 
\cline{2-5} \cline{6-11} 
       &  \textbf{ACC}  & \textbf{PRE} & \textbf{REC} & \textbf{AUC} & \textbf{KAP}                                                             
	 &  \textbf{ACC} & \textbf{PRE}  & \textbf{REC} & \textbf{AUC} & \textbf{KAP} & & \\    \hline
    InceptionV3 & 0.70 & 0.70 & 0.70 & 0.715 & 0.390 & 0.65 & 0.65 & 0.65 & 0.698 & 0.300 & 1,050,114 & 21,802,784 \\\hline
    VGG16 & 0.80 & 0.80 & 0.80 & 0.893 & 0.603 & 0.83 & 0.83 & 0.83 & 0.899 & 0.648 & 263,682 & 14,714,688 \\\hline
    VGG19 & 0.93 & 0.93 & 0.93 & 0.985 & 0.850 & 0.74 & 0.76 & 0.74 & 0.847 & 0.478 & 263,682 & 20,024,384 \\\hline
    ResNet50 & 0.85 & 0.85 & 0.85 & 0.929 & 0.702 & 0.79 & 0.79 & 0.79 & 0.882 & 0.575 & 1,050,114 & 23,587,712 \\\hline
    Xception & 0.72 & 0.78 & 0.72 & 0.741 & 0.425 & 0.65 & 0.70 & 0.65 & 0.698 & 0.316 & 1,050,114 & 20,861,480 \\\hline
    ConvNeXtSmall & 0.84 & 0.85 & 0.84 & 0.933 & 0.676 & 0.85 & 0.85 & 0.85 & 0.937 & 0.693 & 394,754 & 49,454,688 \\\hline
    2D-CNN & 0.95 & 0.95 & 0.95 & 0.984 & 0.903 & 0.96 & 0.96 & 0.96 & 0.991 & 0.910 & 450,914 & 0 \\\hline
    EfficientNet-B0 & 0.96 & 0.96 & 0.96 & 0.998 & 0.918  & 0.91 & 0.91 & 0.91 & 0.970 & 0.820 & 4,010,110 & 42,023 \\
	  \hline \hline
\end{tabular}
\end{table*}

\begin{table}[t]
\centering
\caption{Model performance of the CAE with ML classifiers}
    \begin{tabular}{c|c|c|c|c|c}
        \hline \hline
        \textbf{Models} & \textbf{ACC} & \textbf{PRE} & \textbf{REC} & \textbf{F1-score} & \textbf{AUC}\\\hline
RF & 0.952 & 0.975 & 0.920 & 0.947 & 0.99\\\hline
XGB & 0.948 & 0.951 & 0.936 & 0.944 & 0.99\\\hline
SVC & 0.974 & 0.961 & 0.984 & 0.972 & 1.00\\\hline
KNN & 0.981 & 0.962 & 1.000 & 0.980 & 1.00\\\hline
VC & 0.985 & 0.976 & 0.992 & 0.984 & 1.00\\
        \hline
        \hline
        \end{tabular}
        \label{tab:Time}
\end{table} 


\begin{figure}[!b] 
\centering
\centerline{\includegraphics[width=\linewidth]
{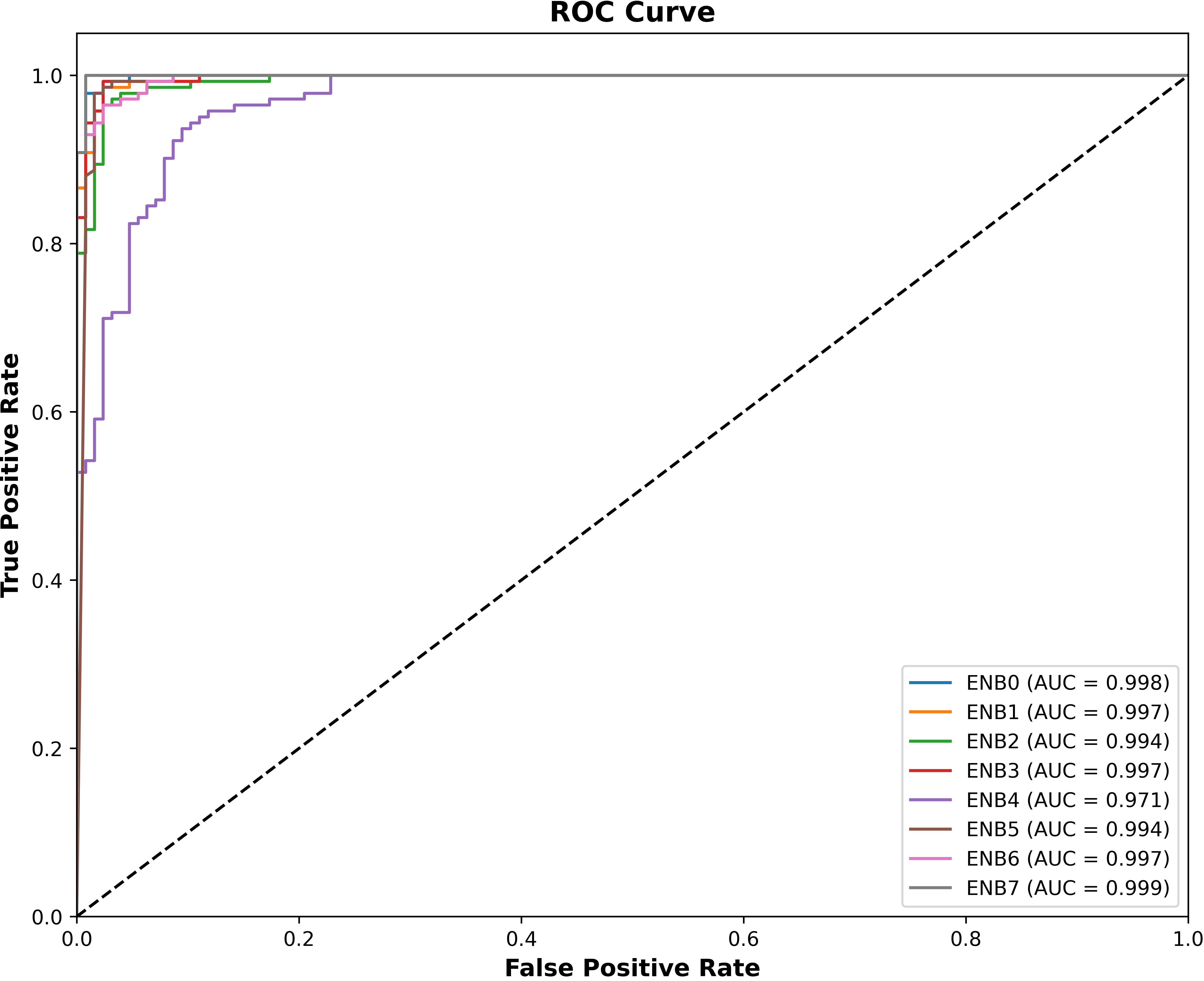}}
\caption{AUC curve of all ENB models combined with SS}
\label{fig}
\end{figure} 

\begin{figure}[!b] 
\centering
\centerline{\includegraphics[width=\linewidth]
{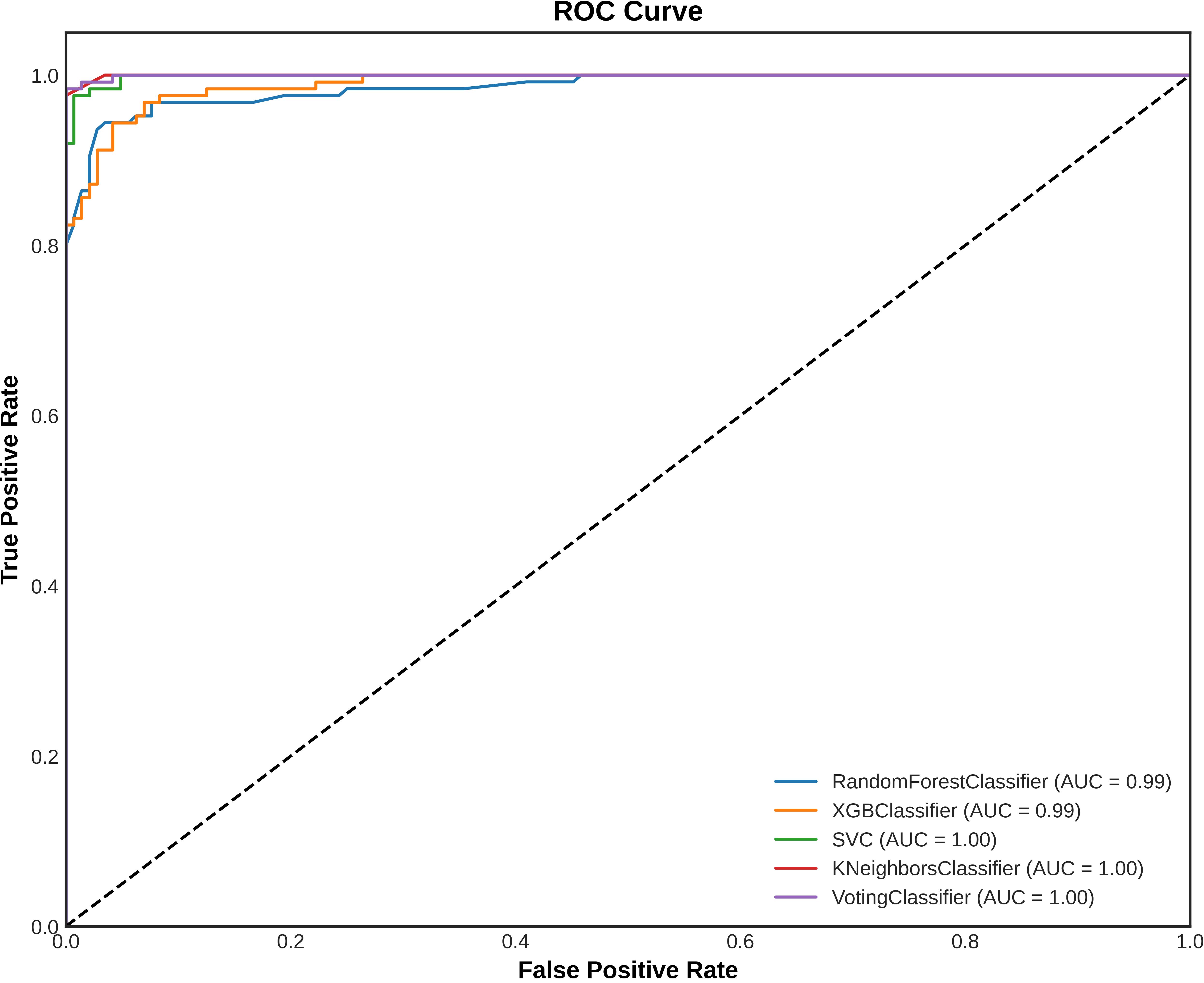}}
\caption{AUC curve of CAE combined with ML models}
\label{fig}
\end{figure} 

\section{RESULTS, DISCUSSION, AND COMPARISON}

\begin{table*}[t!]
\caption{Comparison of our work with other state-of-the-art implementations}
\centering
    \begin{tabular}{c|c|c|c}
        \hline \hline
        \textbf{Author} & \textbf{Method}  & \textbf{EEG dataset}  & \textbf{ACC (\%)}\\
        \hline
        Sui et al. \cite{b40} & MCCA and SVM classifier & 53 HC subjects and 48 SZ patients & 74.00\\
        \hline
        Shim et al. \cite{b51} & BPF with SVM classifier & 34 HC subjects and 34 SZ patients  & 88.24\\
        \hline
        Piryatinska et al. \cite{b41} & $\epsilon$-complexity function with RF classifier & 39 HC subjects and 45 SZ patients  & 84.50\\
        \hline
        Naira et al. \cite{b42} & PCC and CNN & 39 HC subjects and 45 SZ patients  & 90.00\\
        \hline
        Zhang \cite{b43} & ERPs and RF classifier & 32 HC subjects and 49 SZ patients  & 81.10\\
        \hline
        Phang et al. \cite{b44} & TFD features and CNN & 39 HC subjects and 45 SZ patients  & 91.69\\
        \hline
        Siuly et al. \cite{b45} & EMD with EBT & 32 HC subjects and 49 SZ patients & 89.59\\
        \hline
        David et al. \cite{b46} & RCNN & 65 HC subjects and 45 SZ patients  & 89.98\\
        \hline
        Guo et al. \cite{b47}& DNN & 32 HC subjects and 49 SZ patients & 92.00\\
        \hline
        Akbari et al. \cite{b52}& Graphical features and KNN classifier & 14 HC subjects and 14 SZ patients & 94.80\\
        \hline
        Rajesh et al. \cite{b48}& SLBP with Logitboost classifier & 39 HC subjects and 45 SZ patients & 91.66\\
        \hline
        Khare et al. \cite{b49}& SPWVD and AlexNet & 32 HC subjects and 49 SZ patients & 93.33\\ 
        \hline
        & \textbf{CAE with KNN classifier} & & \textbf{98.10}\\
       {\textbf{Proposed Work}}  & \textbf{CAE with VC classifier} & 39 HC subjects and 45 SZ patients & \textbf{98.50}\\
        & \textbf{ SS with 2D-CNN} & & \textbf{95.00}\\
       & \textbf{ SS with EfficientNet-B7} & & \textbf{99.00}\\
        \hline \hline
        \end{tabular}
        \label{tab:Time}     
\end{table*}

In this work, we used EEG data to classify SZ and assessed the effectiveness of CAE and ENB models. The simulation findings provide significant insights into the performance of this approach and its potential for real-time applications in SZ classification. Previous research studies have explored various methodologies for SZ classification. For instance, Sui et al. \cite{b40} developed an algorithm utilizing multi-set canonical correlation analysis (MCCA) and achieved 74\% accuracy with an SVM classifier on EEG data. In another study, Shim et al. \cite{b51} employed a BPF technique to extract features from various frequency bands in EEG data and fed these features to an SVM classifier by obtaining an accuracy of 88.24\%. Piryatinska et al. \cite{b41} proposed an $\epsilon$-complexity function with the objective of creating a low-dimensional feature space while obtaining accuracies of 84.5\% and 81.07\% with RF and SVM classifiers, respectively. Naira et al. \cite{b42} used the Pearson correlation coefficient (PCC) to capture channel relations, reducing the input size (of EEG data) for a CNN and achieving 90\% accuracy in SZ prediction. 

Moreover, Lei Zhang \cite{b43} extracted features from EEG data based on ERP, obtaining the highest accuracy of 81.1\% for SZ classification using an RF classifier. Phang et al. \cite{b44} used time and frequency domain (TFD) features with CNN, achieving the highest accuracy of 91.69\%. Siuly et al. \cite{b45} utilized empirical mode decomposition (EMD) with an ensemble bagged tree (EBT) classifier, resulting in an accuracy of 89.59\%. David et al. \cite{b46} proposed an end-to-end recurrent deep convolutional neural network (RCNN) framework that attained an average accuracy of 89.98\% for SZ detection. On a similar note, Guo et al. \cite{b47} developed a DNN model with an accuracy of 92\% for the classification of SZ from EEG data. In another study, Akbari et al. \cite{b52} employed a two-dimensional phase space dynamic approach to investigate fifteen features' chaotic behavior. They utilized a forward selection algorithm to identify essential features, used a KNN classifier, and achieved an accuracy of 94.8\%. Rajesh et al. \cite{b48} proposed a framework utilizing symmetrically weighted local binary patterns (SLBP) combined with a logistic regression boosting (Logitboost) classifier, achieving an accuracy of 91.66\%. Lastly, our proposed methodology was compared with the smoothed Pseudo-Wigner–Ville distribution (SPWVD) technique combined with TL models \cite{b49}, resulting in an accuracy of 93.34\% using AlexNet and 93.34\% using ResNet50. Apart from SPWVD, they also compared its performance with CWT and STFT and attained the highest accuracy of 90.64\% and 79.18\% using CNN and ResNet50, respectively. 

To assess the efficacy of the proposed approach, we compared it with previously employed methods for SZ classification, including traditional feature extraction methods, CNN, DL, and TL models. The comparison results, shown in Table IV, demonstrate that our model outperforms the aforementioned techniques \cite{b40} - \cite{b49}. The utilization of CAE and ENB models has significantly enhanced the effectiveness of our work. It has enabled us to achieve improved performance when combining both SS and STFT, which would not have been possible solely through conventional DL or TL methods. The incorporation of CAE and ENB models has played a crucial role in elevating the capabilities and outcomes of our study. While our study demonstrates promising results, it is essential to consider the limitations of these models as well. The limitations for ENB models include increased training time, sensitivity to hyperparameter tuning, limited interpretability, domain-specific adaptation requirements, and resource intensiveness for inference. Firstly, It can take a lot of time and computing power to train ENB models, especially for wider and deeper variants like ENB7. Because these models are sophisticated and incorporate a huge number of parameters. Secondly, ENB models are sensitive to hyperparameter tuning. Hyperparameters such as scaling factors, dropout rates, learning rates, and batch sizes require careful tuning to achieve optimal performance. Inadequate tuning can lead to suboptimal results or even training instability. Thirdly, ENB models often have limited interpretability. The complexity and depth of these models make it challenging to understand the learned features and the decision-making process, which can hinder insights and trust in specific applications. Fourthly, ENB models often require fine-tuning or adaptation for optimal performance in domain-specific applications. This adaptation necessitates labelled data specific to the target domain, which may not always be readily available. Lastly, while ENB models are designed to be efficient, larger variants can still demand significant computational resources during inference. This makes real-time applications or deployment on resource-constrained devices challenging. Similarly, CAE also has a few limitations. Firstly, CAEs may struggle to capture temporal dependencies in EEG data, which can be crucial for accurate classification. While they excel at spatial feature extraction, their ability to preserve sequential information across time steps is limited compared to recurrent neural networks. Secondly, the performance of CAEs heavily depends on the availability of large, diverse datasets. Limited or imbalanced data can adversely affect their ability to learn robust and generalizable features, potentially leading to lower performance. Finally, interpreting the features learned by CAEs can be challenging. Understanding the underlying brain mechanisms linked to the SZ condition may be hampered by this lack of interpretability. It is crucial to acknowledge these limitations and carefully evaluate the suitability of CAE and ENB models for specific tasks and resource constraints. Despite these limitations, these models remain powerful tools for various medical applications, offering an appealing balance between performance and efficiency.

\section{CONCLUSION}

In this work, we developed two methods for the classification of SZ from EEG signals employing CAE and ENB models. The proposed approaches offer significant advantages over traditional time-frequency techniques by reducing the need for extensive pre-processing or domain-specific feature extraction of the multi-channel EEG data. The SS generated from the CWT output are fed into various DL and TL models for classification. Additionally, we proposed another approach where decomposed EEG data using DWT was fed into the CAE. After training the CAE model, the encoder part of the CAE was used to extract features, which were then given as input to various ML models for classification. Our approach achieved a high accuracy of 98.5\% using CAE with a VC and 99\% using SS combined with the ENB7 model, making it a reliable and stable method for SZ classification. The CAE model not only serves as a dimensionality reduction technique but also holds potential for real-time or on-chip computing implementations \cite{b80}.

Beyond the current work on developing an automated approach for detecting SZ using the CAE and ENB models, we intend to investigate the application of advanced architectures like Temporal Convolutional Networks (TCNs) and Graph Neural Networks (GNNs) in order to further improve the accuracy and efficiency of the system for feature extraction from EEG signals. TCNs can capture long-range temporal dependencies and hierarchical features, offering an alternative to recurrent architectures for sequential data processing. GNNs, on the other hand, can model the spatial dependencies and interactions across distinct EEG channels, offering a comprehensive understanding of the brain network dynamics in SZ. 

\section{ACKNOWLEDGMENT}
The authors appreciate KongFatt Wong-Lin for helpful comments in an earlier version of the manuscript.

\end{document}